\pdfoutput=1
\let\accentvec\vec % fix for svjour3 conflict with amsmath
\documentclass[smallextended]{svjour3} 

\usepackage[english]{babel}
\usepackage{graphicx}

\let\vec\accentvec
\usepackage{amsfonts,amsmath}
\usepackage{dsfont}

\usepackage{amsfonts,amsmath}

\newcommand{\R}{\ensuremath{\mathds{R}}}
\newcommand{\C}{\ensuremath{\mathds{C}}}
\newcommand{\Cplx}{\ensuremath{\C}}

\newcommand{\ket}[1]{\ensuremath{|#1\rangle}}
\newcommand{\bra}[1]{\ensuremath{\langle#1|}}
\newcommand{\ketbra}[2]{\ensuremath{\ket{#1}\bra{#2}}}
\newcommand{\braket}[2]{\ensuremath{\langle{#1}|{#2}\rangle}}

\newcommand{\1}{{\rm 1\hspace{-0.9mm}l}}
\newcommand{\Id}{\1}

\providecommand{\qed}{\newline\vspace{3mm}\hfill $\Box$}
\providecommand{\proof}{\noindent {\it Proof.\ }}

\newcommand{\tr}{\ensuremath{\mathrm{tr}}}

\newcommand{\ie}{\emph{i.e.}}
\newcommand{\etal}{\emph{et al.}}

\begin{document}

\title{Qubit flip game on a Heisenberg spin chain}
\date{02/08/2011 (v. 0.72)}

\author{J.A. Miszczak\and P. Gawron \and Z. Pucha{\l}a}
\institute{J.A. Miszczak, P. Gawron, Z. Pucha{\l}a \at
Institute of Theoretical and Applied Informatics, Polish Academy
of Sciences, Ba{\l}tycka 5, 44-100 Gliwice, Poland\\\email{miszczak@iitis.pl}}

\maketitle

\begin{abstract}
We study a quantum version of a penny flip game played using control parameters
of the Hamiltonian in the Heisenberg model. Moreover, we extend this game by
introducing auxiliary spins which can be used to alter the behaviour of the
system. We show that a player aware of the complex structure of the system used
to implement the game can use this knowledge to gain higher mean payoff.
\end{abstract}

%%%%%%%%%%%%%%%%%%%%%%%%%%%%%%%%%%%%%%%%%%%%%%%%%%%%%%%%%%%%%%%%%%%%%%%%%%%%%%%%
\section{Introduction}
%%%%%%%%%%%%%%%%%%%%%%%%%%%%%%%%%%%%%%%%%%%%%%%%%%%%%%%%%%%%%%%%%%%%%%%%%%%%%%%%
Game theory is used to describe the situations of mutual interaction of several
parties, where each party aims to maximize its gain. If one considers the
physical system used to define a game, it is reasonable to ask about the
influence of physical laws on games. As a result, one should consider more
general notion of a game, where strategies are defined in terms of quantum
evolution and the system is described in terms of quantum states. In such scheme
the payoff function is defined as a mean value of some chosen observables. As
players in quantum games can use larger space of strategies this should
influence the expected value of the payoff~\cite{meyer99strategies,EWL1999}.

The first example of a quantum game was provided in the form of a simple penny
flip game \cite{meyer99strategies}. In this scenario a player with the ability
to use quantum strategies (moves) can always win with a player using only
classical moves. Eisert \etal\ \cite{EWL1999} provided a more elaborated example
of a two-person quantum game by constructing a quantized version of Prisoner's
dilemma game. Recent developments in this area include the application of
quantum games for the analysis of quantum
walks~\cite{gawron05parrondo,kosik07parrondo,chandrashekar11parrondo} and the
analysis of quantum games in the presence of
decoherence~\cite{flitney02quantum,gawron10noisy,gawron11noise} with the special
attention to the degradation of
entanglement~\cite{gawron08noise,ramazan10distinguishing,ramazan11decoherence}.
Quantum games also provide relatively easy scenarios for implementing quantum
information processing in physical systems~\cite{pakula07mesoscopic} and can be
used to probe the influence of decoherence in such systems~\cite{FA2005}.

The main aim of this work is to provide a scheme for a quantum version of a
penny flip game, based on the Heisenberg chain model. We study a protocol for
playing a quantum game where two players implement their moves using control
parameters of the model. We show that for a two-qubit spin chain it is possible
to harness the complex structure of the system to obtain a higher mean payoff. 

This paper is organized as follows.
In Section~\ref{sec:penny-flip} we review a~classical penny flip (or bit flip)
game including a quantum versus classical player scenario.
In Section~\ref{sec:qubit-flip} we discuss the quantized version of the penny
flip game played using control parameters of a single qubit.
In Section \ref{sec:control} we extend the previous model of the penny flip game
and discuss quantum strategies which can be achieved using an auxillary qubit.
In Section~\ref{sec:final} we provide the summary of the presented work and give
some concluding remarks.

%%%%%%%%%%%%%%%%%%%%%%%%%%%%%%%%%%%%%%%%%%%%%%%%%%%%%%%%%%%%%%%%%%%%%%%%%%%%%%%%
\section{Penny flip game}\label{sec:penny-flip}
%%%%%%%%%%%%%%%%%%%%%%%%%%%%%%%%%%%%%%%%%%%%%%%%%%%%%%%%%%%%%%%%%%%%%%%%%%%%%%%%
Penny flip game, also known as a bit flip game, provides one of the simplest 
examples used in many textbooks on game theory. It can be also easily used
to show that by using quantum strategies, represented by unitary gates, one of
the players can always win. 

%%%%%%%%%%%%%%%%%%%%%%%%%%%%%%%%%%%%%%%%%%%%%%%%%%%%%%%%%%%%%%%%%%%%%%%%%%%%%%%%
\subsection{Classical version}
%%%%%%%%%%%%%%%%%%%%%%%%%%%%%%%%%%%%%%%%%%%%%%%%%%%%%%%%%%%%%%%%%%%%%%%%%%%%%%%%
Let us consider two parties (players) -- Alice and Bob -- playing a strategic
game consisting of flipping a coin. At the beginning of the game the coin is
faced heads up. The game consists of three moves. Alice plays first, then Bob
and then Alice again. Each of them can chose to flip or not to flip the coin.
The coin is hidden, so they cannot see it. Each of them does not know the
actions of the other. The goal of Alice is that the coin ends facing tails up.
The goal of Bob is that it ends heads up.

The payoff matrix for the penny-flip game is presented in
Table~\ref{tab:penny-flip}, where N is used to denote the not-flipping strategy
and F denotes the flipping strategy. One means that Alice has won, minus one
means that Bob has won and the game is a zero-sum game.

\begin{table}[ht]
\begin{center}
\begin{tabular}{|c|c|c|c|c|}
\hline
\ & NN & NF & FN & FF \\ 
\hline
N & -1 & 1 & 1 & -1 \\ 
\hline
F & 1 & -1 & -1 & 1 \\
\hline
\end{tabular}
\end{center}
\caption{Payoff matrix for the penny flip game.}
\label{tab:penny-flip} 
\end{table}

The above scenario can be described mathematically in a language of quantum
mechanics by using state vectors and unitary gates. In such description a coin
is replaced by a qubit and we map a coin facing heads up to state $\ket{0}$, and
facing tails up to $\ket{1}$. The flip operation corresponds to unitary matrix 
$F=\sigma_x=\left(
\begin{smallmatrix}
0 & 1 \\ 
1 & 0
\end{smallmatrix}
\right)
$ 
\ie\ quantum $Not$ gate, while not-flipping operator is described by the
identity matrix,
$N=\1=
(\begin{smallmatrix}
1 & 0 \\ 
0 & 1
\end{smallmatrix}).
$

The probability of winning for Alice and Bob can be calculated as the
expectation value of the $\sigma_z$ operator 
\begin{equation}
\langle\sigma_z\rangle_{\ket{c}}=\tr(\ketbra{c}{c}\sigma_z),
\end{equation}
where $\ket{c}$ is the state of the coin after the second move of Alice.

One can easily see that in this scenario there is no winning strategy neither
for Alice nor for Bob.

%%%%%%%%%%%%%%%%%%%%%%%%%%%%%%%%%%%%%%%%%%%%%%%%%%%%%%%%%%%%%%%%%%%%%%%%%%%%%%%%
\subsection{Quantum versus classical player}\label{sec:quantum-vs-classical}
%%%%%%%%%%%%%%%%%%%%%%%%%%%%%%%%%%%%%%%%%%%%%%%%%%%%%%%%%%%%%%%%%%%%%%%%%%%%%%%%
In order to show how quantum mechanics can be used to cheat in the penny-flip
game one should note that in the above description Alice and Bob have only a
very limited set of operations at their disposal.

Let us now use a qubit instead of a penny and allow one of the players, namely
Alice, to use any unitary operation during her moves. In this situation any
normalised linear combination of the base states, $\alpha\ket{0}+\beta\ket{1},
\alpha,\beta\in\Cplx$ with $|\alpha|^2+|\beta|^2=1$, describes an allowed state
of a qubit representing a coin. Let us also assume that only Alice knows that
the game is played with a qubit. 

The above condition allows Alice to win with probability 1. She can rotate the
qubit in any direction and prepare the state which is invariant with respect to
any move Bob can use. To achieve this, she may chose to apply a Hadamard gate, 
$H=\frac{1}{\sqrt{2}} 
\left(\begin{smallmatrix}
  1 & 1 \\
  1 & -1
\end{smallmatrix}
\right),$
during her both moves. Bob still has only two possibilities: to flip
($\sigma_x$) or
not to flip ($\1$). In these two cases we have
\begin{enumerate}
 \item $H \sigma_x H\ket{0} \rightarrow \ket{0},$
 \item $H \1 H\ket{0} \rightarrow \ket{0}.$
\end{enumerate}
In the last step Alice can prepare a coin in state $\ket{1}$ by using $\sigma_x$
operator. From the above it follows that Bob's actions have no influence
whatsoever on the final state of the coin and thus the outcome of the game. 

One should note that we have restricted ourselves to pure strategies. More
detailed analysis of this game including mixed (probabilistic) strategies may be
found in~\cite{PS2003}.

%%%%%%%%%%%%%%%%%%%%%%%%%%%%%%%%%%%%%%%%%%%%%%%%%%%%%%%%%%%%%%%%%%%%%%%%%%%%%%%%
\section{Qubit flip game}\label{sec:qubit-flip}
%%%%%%%%%%%%%%%%%%%%%%%%%%%%%%%%%%%%%%%%%%%%%%%%%%%%%%%%%%%%%%%%%%%%%%%%%%%%%%%%
In the scheme described in Section~\ref{sec:quantum-vs-classical} the player
using quantum strategies was able to always win. This situation was possible
since the player using classical strategies was not able to explore the possible
space of states. As such, it is natural to ask how the situation changes if both
players are able to use quantum moves.

Let us assume that both Alice and Bob are now aware of the fact that they are 
playing the game using a qubit. Therefore now we have the following game
\begin{equation}
  \Gamma=(\mathcal{H},\ket{\psi}, S_A, S_B, P_A, P_B),
\end{equation}
where $\mathcal{H}=\Cplx^2$ is a Hilbert space, $\ket{\psi}=\ket{0}$ is an
initial state, $S_A=U(2)\times U(2)$ the set of Alice's strategies, $S_B=U(2)$
the set of Bob's strategies and $P_A$, $P_B$ are respectively Alice's and Bob's
payoffs which are defined below.

The course of the game is as follows. Alice plays first -- she applies her first
element of strategy, $U_{A_1}$, on the initial state, then Bob plays applying
his strategy $U_B$, finally Alice plays by applying the second element of her
strategy, $U_{A_2}$. Then the projective measurement
$\{O_\mathrm{heads}\rightarrow\ketbra{0}{0},
O_\mathrm{tails}\rightarrow\ketbra{1}{1}\}$ is performed. The final state of the
game (before measurement) is 
\begin{equation}\label{equ:game}
  \ket{\psi_f}=U_{A_2} U_{B} U_{A_1}\ket{0}.
\end{equation}

Alice's and Bob's payoff functions depend on the probability of measuring
$O_\mathrm{tails}$ and $O_\mathrm{heads}$ respectively
$P_A(S_A,S_B)=|\braket{1}{\psi_f}|^2$,
$P_B(S_A,S_B)=|\braket{0}{\psi_f}|^2=1-P_A(S_A,S_B)$.

%%%%%%%%%%%%%%%%%%%%%%%%%%%%%%%%%%%%%%%%%%%%%%%%%%%%%%%%%%%%%%%%%%%%%%%%%%%%%%%%
\subsection{Two quantum players with one qubit}
%%%%%%%%%%%%%%%%%%%%%%%%%%%%%%%%%%%%%%%%%%%%%%%%%%%%%%%%%%%%%%%%%%%%%%%%%%%%%%%%
Pure strategies in this game cannot be in Nash equilibrium, that is why 
in searching for optimality the players will use mixed strategies.
To obtain Nash equilibrium the first or the second player can use the variety of 
mixed strategies. Here we give examples of two such strategies, namely 
\emph{Pauli strategy} and \emph{Haar strategy}. We will show that these 
strategies give Nash equilibrium.
\begin{itemize}
    \item \emph{Pauli strategy} is a mixed strategy, where a player chooses one
    of four unitary matrices $\{ \1, i \sigma_x, i \sigma_y, i \sigma_z \}$ with
    equal probability.
    \item In \emph{Haar strategy} a player chooses a random special unitary
    matrix distributed with natural probability measure invariant with respect
    to unitary transformation, \ie{} the Haar measure.
\end{itemize}

We will show that if the second player uses one of the above strategies, then,
no matter what action the first player will take, the probabilities of the
success for both players are equal. We start with a definition. 

\begin{definition}
A finite subset $X$ of $U(d)$ is unitary $t$-design \cite{roy09unitary} if 
\begin{equation}
\frac{1}{|X|}\sum_{U\in X} U^{\otimes t}\otimes (U^\star)^{\otimes t} = 
\int_{U(d)} U^{\otimes t}\otimes (U^\star)^{\otimes t} d\mu (U),
\end{equation}
where $\mu$ is the Haar measure on a group $U(d)$.
\end{definition}

\begin{lemma}
Let $X$ be a unitary 1-design. Then a mixed strategy such that the second player
chooses to play a random element $U \in X$ chosen with equal probability, gives
Nash equilibrium.
\end{lemma}
\proof 
Let us denote the unitary matrices played by the first player as
\begin{eqnarray}
 U_{A_1} = \{a_{ij}\}_{i,j=1}^2 , \quad
 U_{A_2} = \{b_{ij}\}_{i,j=1}^2.
\end{eqnarray}
Then the probability of success for the second player is given by
\begin{equation}
 \bar{P}_B = \int_{U(2)} \left| \bra{0}U_{A_2} U U_{A_1}\ket{0}\right|^2  d
\mu_X(U) ,
\end{equation}
where $\mu_X$ is a uniform probability measure supported on $X$.
% 
% In the case of Pauli strategy $\mu$ is discrete uniform measure supported on 
% $\{ \1, i \sigma_x, i \sigma_y, i \sigma_z \}$.
%Pauli matrices plus identity. 
% In the case of Haar strategy measure $\mu$ is the Haar measure on 
% a group of special unitary matrices $SU(2)$.
Simple calculations give us 
\begin{eqnarray}
\bar{P}_B&=&\int_{U(2)} \left|\bra{0}U_{A_2} U U_{A_1}\ket{0} \right|^2 d
\mu_X(U) \\
%&&\int \left|\sum_{i,j=1}^2 u_{ij} a_{1i} b_{j1}\right|^2 d \mu(U) = \\
&=& \sum_{i,j=1}^2 |a_{1i} b_{j1}|^2 \int_{U(2)} |U_{ij}|^2 d \mu_X(U) \\
&&+ 
\sum_{\substack{i \neq k \\ j \neq l}} a_{1i} b_{j1} \overline{a_{ki} b_{l1}} \int_{U(2)} U_{ij} \overline{U_{kl}}
\ d \mu_X(U).
\end{eqnarray}
Since $X$ is a unitary 1-design, the values of the above integrals are the same
as the value of integrals with respect to Haar measure
\cite{collins06integration}
\begin{equation}
\int_{U(2)} U_{ij}\overline{U_{kl}} d\mu(U) = \frac{1}{2}
\delta_{ik}\delta_{j,l},
\end{equation}
where $\mu$ is Haar measure on a group $U(2)$.
Finally, since $U_{A_1}, U_{A_2}$ are unitary, we get
\begin{eqnarray}
\bar{P}_B = \frac{1}{2} \sum_{i,j=1}^2 |a_{1i} b_{j1}|^2 = \frac{1}{2}.
\end{eqnarray}
Thus, if the second player chooses to play a strategy given by a 1-design, no
player can benefit by altering their strategy.
\qed

Since the Pauli strategy forms a unitary 1-design, one can see that the Pauli
strategy gives Nash equilibrium.

The same holds true for the Haar strategy, but in this case to obtain
the result one needs to notice that 
\begin{equation}
\int_{U(2)} U\otimes U^\star d\mu_{S} (U) = 
\int_{U(2)} U\otimes U^\star d\mu (U),
\end{equation}
where $\mu_{S}$ is a Haar measure on a group $SU(2)$.

From the above considerations one can see that to achieve Nash equilibrium, Bob
can take any mixed strategy, in which he chooses unitary matrices with the
distribution $\nu$ satisfying the condition 
\begin{equation}
\int_{U(2)} U \otimes U^\star d\nu(U) = \int_{U(2)} U \otimes U^\star d\mu(U).
\end{equation}
Moreover, it is easy to see the following.
\begin{proposition}
If the second player uses a mixed strategy by choosing uniformly from a finite
set $X$, then this strategy gives a Nash equilibrium if and only if set $X$ is a
unitary 1-design.
\end{proposition}

The above result suggests that, if Bob aims to play optimally, he should always
play a strategy given be a unitary 1-design. In the following we assume that he
always uses the Pauli strategy.

%%%%%%%%%%%%%%%%%%%%%%%%%%%%%%%%%%%%%%%%%%%%%%%%%%%%%%%%%%%%%%%%%%%%%%%%%%%%%%%%
\subsection{Hamiltonian-based implementation on one qubit}
%%%%%%%%%%%%%%%%%%%%%%%%%%%%%%%%%%%%%%%%%%%%%%%%%%%%%%%%%%%%%%%%%%%%%%%%%%%%%%%%
The rules of the qubit-flip game can be described to reflect the constraints
that are imposed upon the physical realizations of the game. The players will be
allowed to program a~quantum device that represents a one-qubit system. They
will be constrained by the execution time of the program and the set of
available instructions.
% 
% In the following we denote by $\mathrm{SU}(d)$ the set of special unitary
%matrices 
% of dimension $d$.

Let us assume that, instead of being able to execute a fixed unitary
transformation, Alice and Bob are now allowed to play the bit-flip game by
manipulating the control parameters in the Hamiltonian $H=H_c(t)$. Note that
since there is no drift term in this equation, we assume that players have
complete control over the system. They are only constrained by maximal fixed
time $T$ of the game turn. During this time Alice and Bob can change values of
the parameters $h_z$ and $h_y$.

The realisation of the game in this set-up is as follows. Alice moves as the
first player and applies the series of control to the system, then Bob performs
his sequence of controls and, finally, Alice performs her second sequence as the
last move in the game. Each move is described by a sequence of parameters
$
h_y(0), h_z(\Delta t), h_y(2\Delta t), \ldots, h_i((N-1) \Delta t),
$
with $h_i=h_y$ or $h_i=h_z$ depending on the length of the sequence. We assume
that players cannot use $h_z$ and $h_y$ simultaneously. We will only consider
the case where the time for which the control is applied is equal for all the
controls.

The final state of the game can be obtained from Eq.~(\ref{equ:game}) with
\begin{eqnarray}
U_{A_1}&=&
e^{-i h_i^{A_1}((N_{A_1}-1)\Delta t)\sigma_i}
\cdots
e^{-i h_z^{A_1}(\Delta t)\sigma_z}
e^{-i h_y^{A_1}(0)\sigma_y},
\\
U_{B}&=&
e^{-i h_i^{B}((N_{B}-1)\Delta t)\sigma_i}
\cdots
e^{-i h_z^{B}(\Delta t)\sigma_z}
e^{-i h_y^{B}(0)\sigma_y}
,\\
U_{A_2}&=&
e^{-i h_i^{A_2}((N_{A_2}-1)\Delta t)\sigma_i}
\cdots
e^{-i h_z^{A_2}(\Delta t)\sigma_z}
e^{-i h_y^{A_2}(0)\sigma_y}.
\end{eqnarray}

To obtain the Pauli or Haar strategy the second player may calculate control
sequences using Euler decomposition of $SU(2)$. For a given special unitary
matrix one can write
\begin{equation}
\left(
%\begin{array}{cc}
\begin{smallmatrix}
 e^{i \phi } \cos (\theta ) & e^{i \psi } \sin (\theta ) \\
 -e^{-i \psi } \sin (\theta ) & e^{-i \phi } \cos (\theta )
\end{smallmatrix}
%\end{array}
\right) = 
e^{i \frac{\phi + \psi}{2}  \sigma_z} 
e^{i \theta \sigma_y} 
e^{i \frac{\phi - \psi}{2}  \sigma_z}.
\end{equation}
Thus to obtain the Pauli strategy the second player chooses uniformly one of
four matrices from the Pauli strategy and takes control parameters $(\xi_1,
\xi_2, \xi_3)$ from Table~\ref{tab:control-pauli}.

\begin{table}
\begin{center}
  \begin{tabular}{| c | c | c | c |}
    \hline
         	&    $\xi_1$	&    $\xi_2$	&    $\xi_3$\\ \hline
 $\Id$ 	& 0	& 0	& 0\\ \hline
 $i \sigma_x$      	& $ \pi/4$  	& $-  \pi/2$	& $ \pi/4$\\ \hline
 $i \sigma_y$    	&0	& $-\pi/2$	&0\\ \hline
 $i \sigma_z$     	& $-\pi/4$ 	& $0$ 	& $-\pi/4$ \\ \hline
  \end{tabular}
\end{center}
\caption{Control parameters for realizing the Pauli strategy~\cite{DAlessandro2008introduction}.}
\label{tab:control-pauli}
\end{table}

The player applies parameters as a control sequence with constant factor
$\frac{3}{T}$, each of them for one-third of total time $T$, obtaining 
\begin{equation}
e^{-i \xi_3 \sigma_z} 
e^{-i \xi_2 \sigma_y} 
e^{-i \xi_1 \sigma_z},
\end{equation}
which is one of four matrices in the Pauli strategy.

On the other hand, to obtain the Haar strategy the second player may choose
parameters $\xi_1, \xi_2, \xi_3$ as 
\begin{equation}
 \xi_1 = - \frac{\phi - \psi}{2},\quad 
 \xi_2 = - \arcsin(\sqrt{p}),\quad
 \xi_3 = - \frac{\phi + \psi}{2},
\end{equation}
where $\phi, \psi, p$ are independent random variables with $\phi, \psi \sim
\mathrm{U}(0,2 \pi)$ and $p \sim \mathrm{U}(0,1)$. The player applies the chosen parameters with
constant factor $\frac{3}{T}$, each of them for one-third of total time $T$,
obtaining a random special unitary matrix with Haar distribution~\cite{ZK94}
\ie\ matrix from the Haar strategy.

%%%%%%%%%%%%%%%%%%%%%%%%%%%%%%%%%%%%%%%%%%%%%%%%%%%%%%%%%%%%%%%%%%%%%%%%%%%%%%%%
\section{Cheating with auxiliary spins}\label{sec:control}
%%%%%%%%%%%%%%%%%%%%%%%%%%%%%%%%%%%%%%%%%%%%%%%%%%%%%%%%%%%%%%%%%%%%%%%%%%%%%%%%
% For a given two matrices $U,U_\mathrm{target}\in\mathrm{SU}(d)$ we define
% \emph{gate fidelity} $F\in[0,1]$ of $U$ with respect to $U_\mathrm{target}$ as
% \begin{equation}
% F=\frac{1}{d}\left|\tr[ U^\dagger U_\mathrm{target}]\right|.
% \end{equation}

Let us now consider an extension of the qubit-flip scenario where one of the
players, namely Alice, is aware of the fact that the system used to play the
game is composed of two qubits. In this situation one can ask if, analogously to
the situation in classical versus quantum player, she can use this knowledge to
get a better mean payoff.

%%%%%%%%%%%%%%%%%%%%%%%%%%%%%%%%%%%%%%%%%%%%%%%%%%%%%%%%%%%%%%%%%%%%%%%%%%%%%%%%
\subsection{Description of the scheme}
%%%%%%%%%%%%%%%%%%%%%%%%%%%%%%%%%%%%%%%%%%%%%%%%%%%%%%%%%%%%%%%%%%%%%%%%%%%%%%%%
In the following we are interested in a specific class of systems where the
Hamiltonian operator of the system is given as
\begin{equation}\label{eqn:hamiltonian}
H(t)=H_0+H_c(t)
\end{equation}
where $H_0$ is the drift Hamiltonian and $H_c$ is the control Hamiltonian.

In the case of time-independent Hamiltonian $(H_c(t)\equiv 0)$ the unitary
matrix $U$ describing the time evolution of the system in question is obtained
from the Hamiltonian operator of the system by $U = \exp\left(-i t H_0\right)$
for a given $t$.

However, we are interested in the situation where $H_c(t)\not\equiv 0$ and we
are going to consider the control Hamiltonian, such that control parameters are
piecewise constant in time. In this situation the resulting unitary evolution is
of the form
\begin{equation}\label{eqn:unitary-by-control}
U=\prod_{n=0}^{N-1} e^{-i\Delta t(H_0+H_c(n\Delta t))}
%&=&e^{-i\Delta t(H_0+H_c((N-1)\Delta t))}\ldots e^{-i\Delta t(H_0+H_c(0))}\\
=U((N-1)\Delta t)\ldots U(\Delta t)U(0).
\end{equation}
The Hamiltonian operator of the system used to play the game in such case is 
given as in Eq.~(\ref{eqn:hamiltonian}) with
\begin{equation}
H_0= J
(\sigma_{x}\otimes\sigma_{x}+\sigma_{y}\otimes\sigma_{y}+\sigma_{z}\otimes\sigma_{z})
\end{equation}
and
\begin{equation}
H_c(t)= h_y(t)\sigma_{y}\otimes \Id + h_z(t)\sigma_{z}\otimes \Id.
\end{equation}
Here $\sigma_x,\sigma_y,\sigma_z$ are Pauli matrices
$
\sigma_{x}=(\begin{smallmatrix}
0&1\\
1&0
\end{smallmatrix}),
\sigma_{y}=(\begin{smallmatrix}
0&-i\\
i&0
\end{smallmatrix})
,
\sigma_{z}=(\begin{smallmatrix}
1&0\\
0&-1
\end{smallmatrix})
$
and the parameter $J\in \R$ is the coupling constant~\cite{HBBS2010}.
% 
% In this situation $U_\mathrm{target}$ is the desired evolution of the system,
% whilst 
% $$
% U=U((N-1)\Delta t)\ldots U(\Delta t)U(0)\in\mathrm{SU}(4)
% $$
% describes the evolution resulting from the use of a sequence $\mathbf{h}$ of
% control parameters
% $$
% h_y(0), h_z(\Delta t), h_y(2 \Delta t), \ldots, h_y((N-2) \Delta t), h_z((N-1)
%\Delta t).
% $$
% Note that control are interlaced. 
% 
% Our aim is to maximize the gate fidelity of $U$ wrt some given unitary matrix
% $U_\mathrm{target}$.
% 
% Let us assume that $J=1$. For a given $U_\mathrm{target}$, number of intervals
%$N$
% and control time $\Delta t$ our aim is to find $\mathbf{h}$ such that
% \begin{equation}
% F = \frac{1}{4}\left|\tr U^\dagger U_\mathrm{target} \right|
% \end{equation}
% is maximal.

%%%%%%%%%%%%%%%%%%%%%%%%%%%%%%%%%%%%%%%%%%%%%%%%%%%%%%%%%%%%%%%%%%%%%%%%%%%%%%%%
\subsection{Mimicking one-spin behaviour}
%%%%%%%%%%%%%%%%%%%%%%%%%%%%%%%%%%%%%%%%%%%%%%%%%%%%%%%%%%%%%%%%%%%%%%%%%%%%%%%%
If Alice aims to cheat against Bob by using auxiliary spins, the first challenge
she has to face is to convince Bob that the system is constructed in such a way
that both players can win with equal probability. Let us assume that Bob can
study the system used to play the game with a series of experiments before
actually playing the game. The only possible experiment Bob can make is to play the
game against Alice with the best possible strategy and observe his mean
payoff. If the payoff is equal to $1/2$, then he must assume that both players
have an equal probability of wining.

In order to achieve this effect, Alice must choose the value of $J/T$ in such
way, that if both players use the Pauli strategy they cannot distinguish if they
are using a device composed of one or two qubits using the information about the
mean payoff only.

Let us assume that $T=1$. If the system (the spin chain) used to play the game
is composed of two qubits only, Alice is able to convince Bob that they are
using one qubit only.

\begin{figure}[htp!]
\includegraphics[width=0.95\columnwidth]{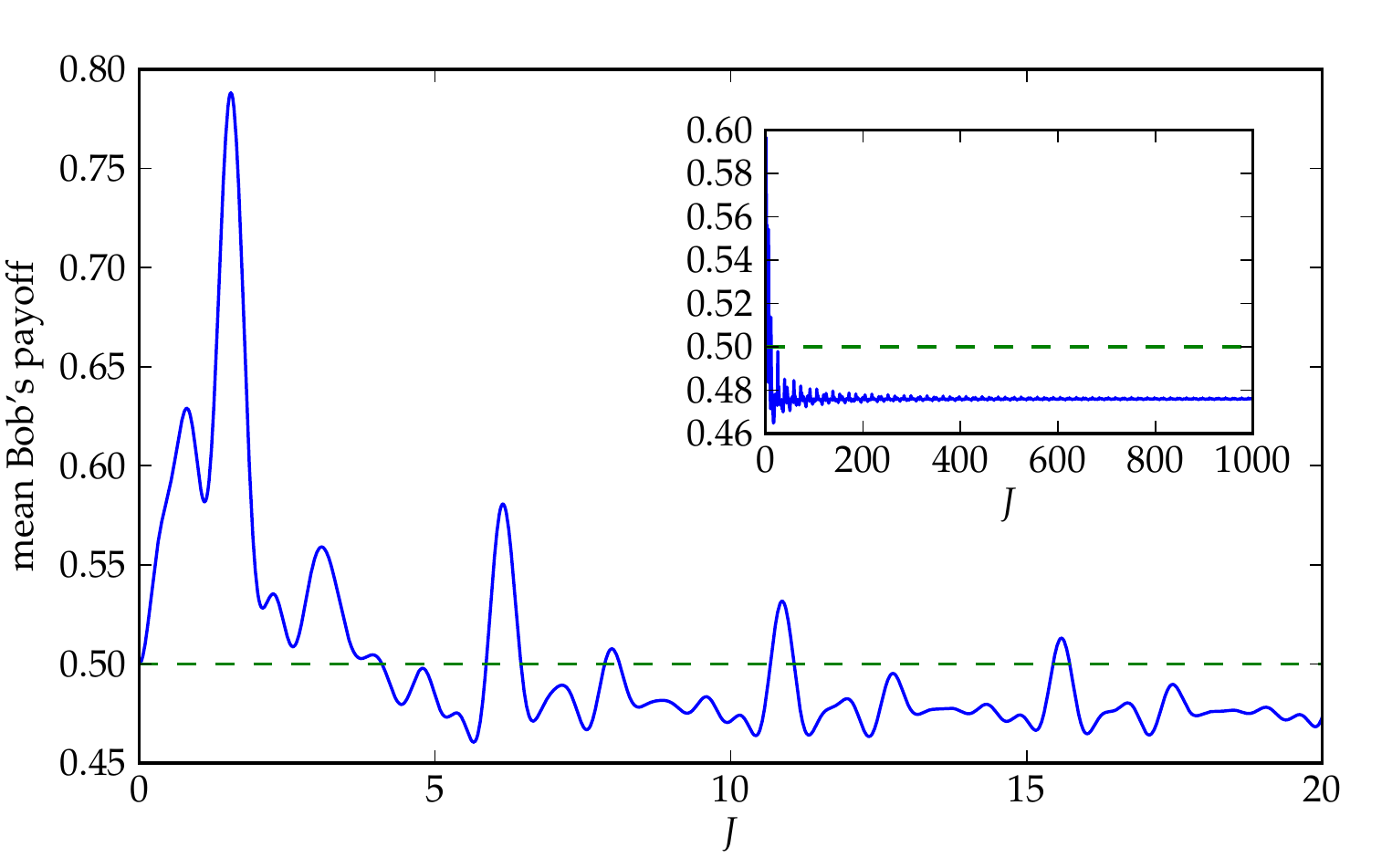}
\caption{Mean value of Bob's payoff in function of spin chain coupling strength
$J$ when both players use Pauli strategy. The inset plot shows the behaviour of
the payoff for very large values of $J$.}
\label{fig:j2}
\end{figure}

In Fig.~\ref{fig:j2} Bob's mean payoffs for different values of the coupling
constant $J$ are presented. In the case of a spin chain composed of two spins, the
smallest value of $J$ greater than zero, such that the condition for the game
appears to be fair, is equal to $J/T\approx 4.10469$.

%%%%%%%%%%%%%%%%%%%%%%%%%%%%%%%%%%%%%%%%%%%%%%%%%%%%%%%%%%%%%%%%%%%%%%%%%%%%%%%%
\subsection{Mean payoff for Alice}
%%%%%%%%%%%%%%%%%%%%%%%%%%%%%%%%%%%%%%%%%%%%%%%%%%%%%%%%%%%%%%%%%%%%%%%%%%%%%%%%
Having the value of $J/T$ fixed, we can assume that Bob always uses the
optimal strategy, \ie\ he aims to achieve the probability of wining equal to $1/2$.
In the following we assume that he always plays the Pauli strategy and applies 
three controls.

Let be given $H_0$, the drift Hamiltonian, which is known to Alice only and $T$,
the time for each move, known to both players. Alice's goal is to optimize
the following functional 
\begin{equation}
P_A(\mathbf{h}_1,\mathbf{h}_2)=\frac{1}{4}\sum_{i=1}^{4} \bra{1} \tr_2
(\ketbra{\psi_i^f}{\psi_i^f})\ket{1}
\end{equation}
with respect to control vectors $\mathbf{h}_1$ and $\mathbf{h}_2$. The final
state of the game $\ket{\psi_i^f}$ reads 
\begin{equation}
\ket{\psi_i^f}=U(\mathbf{h}_2) V_i U(\mathbf{h}_1) \ket{00},
\end{equation}
where matrices $U(\mathbf{h}_1)$ and $U(\mathbf{h}_2)$ are defined as in Eq.~(\ref{eqn:unitary-by-control}).
Matrices $V_i$ describe one of Bob's possible moves, 
\begin{equation}
V_i=e^{-i\frac{T}{3}(H_0+\frac{3}{T}\xi_3^{(i)}\sigma_z\otimes\1)}
e^{-i\frac{T}{3}(H_0+\frac{3}{T}\xi_2^{(i)}\sigma_y\otimes\1)}
e^{-i\frac{T}{3}(H_0+\frac{3}{T}\xi_1^{(i)}\sigma_z\otimes\1)},
\end{equation}
where $\xi_k^{(i)}, k=1,2,3$  are control parameters from $i$-th row of
Table~\ref{tab:control-pauli}.

%%%%%%%%%%%%%%%%%%%%%%%%%%%%%%%%%%%%%%%%%%%%%%%%%%%%%%%%%%%%%%%%%%%%%%%%%%%%%%%%
\subsubsection{Optimal strategy for Alice}
%%%%%%%%%%%%%%%%%%%%%%%%%%%%%%%%%%%%%%%%%%%%%%%%%%%%%%%%%%%%%%%%%%%%%%%%%%%%%%%%
In order to find optimal control sequences allowing to maximize the final output
we have used nonlinear conjugate gradient algorithm \cite{wright99numerical}
implemented as a part of SciPy open-source mathematical software distributed
with modified BSD license \cite{scipy}.

\begin{figure}[ht!]
\centering
\includegraphics[width=\textwidth]{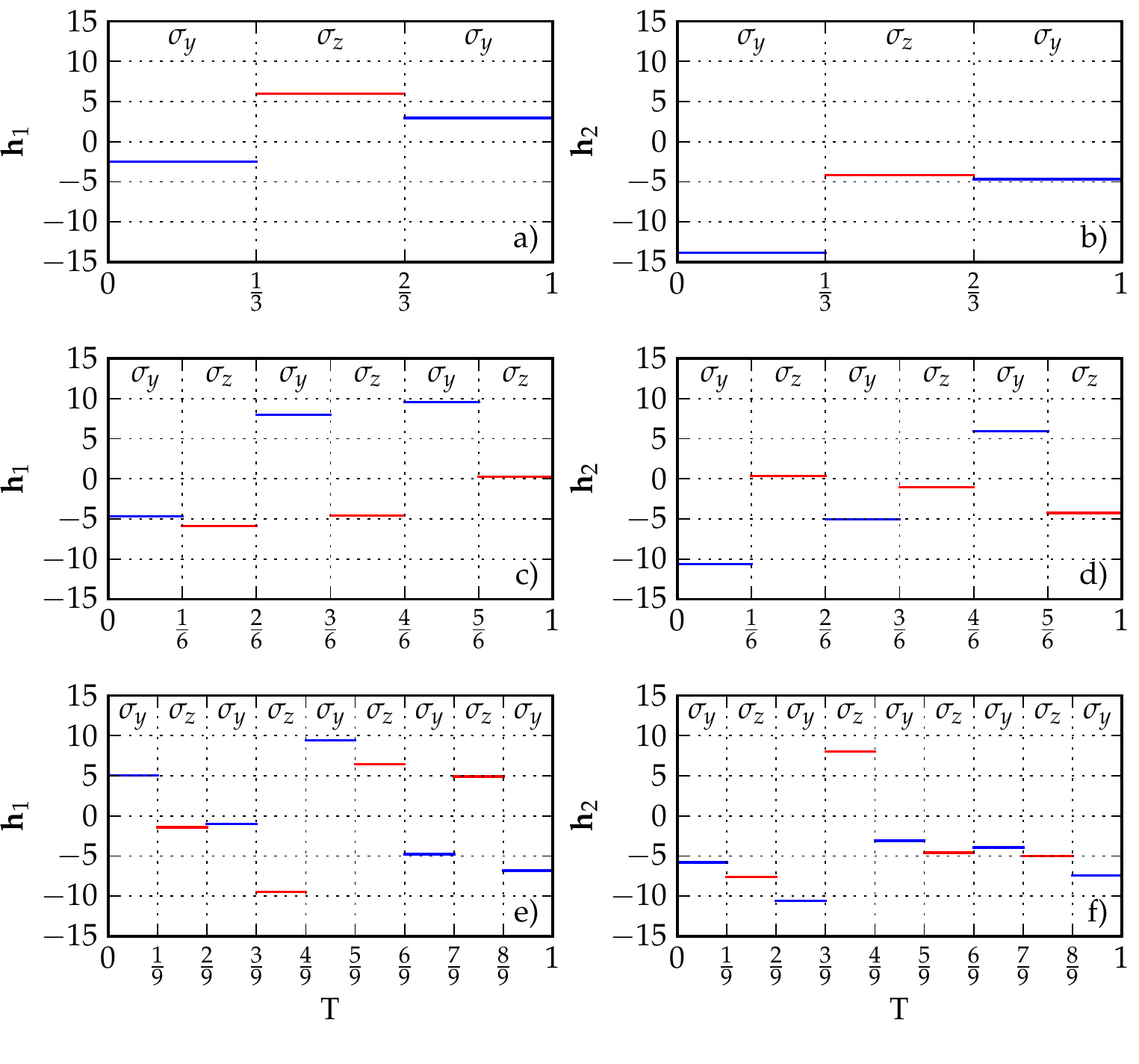}
\caption{Controls that maximize Alice's payoff. Panels a) and b) show the sequence
of pulses that generate gates $U_{A_1}$ and $U_{A_2}$ respectively when Alice
chooses to apply only three controls to run her strategy. In this case her
probability of winning is equal to $0.97$. Similarly panels c) and d) represent
her strategy using 6 pulses per gate; in this case her probability of winning is
$0.988$. Panels e) and f) represent 9-pulse strategy which gives the probability
of winning $0.999$.}
\label{fig:controlsAlice}
\end{figure}

In Fig.~\ref{fig:controlsAlice} the sequences of control pulses allowing Alice to
obtain mean payoff equal approximately to 1 are presented. It is easy to observe
that she is able to exploit the internal structure of the system by using a very
short sequence of control pulses. Additionally, the gain obtained by increasing 
the length of control sequences is very small. 

From the Fig.~\ref{fig:controlsAlice} one can conclude that the knowledge of the
dimensionality of the system can be easily used to cheat against the opponent.
This situation resembles the one described in
Sec.~\ref{sec:quantum-vs-classical}, where one of the players was playing using
the classical moves only. In our case, however, both players are able to exploit
quantum strategies.

%%%%%%%%%%%%%%%%%%%%%%%%%%%%%%%%%%%%%%%%%%%%%%%%%%%%%%%%%%%%%%%%%%%%%%%%%%%%%%%%
\subsubsection{Optimal strategy for Bob}
%%%%%%%%%%%%%%%%%%%%%%%%%%%%%%%%%%%%%%%%%%%%%%%%%%%%%%%%%%%%%%%%%%%%%%%%%%%%%%%%
Let us now assume that the device used to play the game was prepared not by
Alice, but by Bob. In such situation only Bob is aware of the internal structure
of the system and he may use it to increase his payoff.

We can distinguish the following two scenarios where Bob aims to cheat against 
Alice:
\begin{enumerate}
	\item Bob uses one auxiliary qubit and he tries to find the best
	possible sequence of controls;
	\item Bob adds more auxiliary qubits and tries to find the best sequence of
	control pulses.
\end{enumerate}

%%%%%%%%%%%%%%%%%%%%%%%%%%%%%%%%%%%%%%%%%%%%%%%%%%%%%%%%%%%%%%%%%%%%%%%%%%%%%%%%
\paragraph{Optimization over the number of controls}
%%%%%%%%%%%%%%%%%%%%%%%%%%%%%%%%%%%%%%%%%%%%%%%%%%%%%%%%%%%%%%%%%%%%%%%%%%%%%%%%
We may assume that now Alice is using the Pauli strategy. 
The $J/T$ was chosen so that both players 
have no reasons for suspecting that the system has higher dimensionality. 

\begin{figure}[ht!]
\centering
\includegraphics[width=0.95\textwidth]{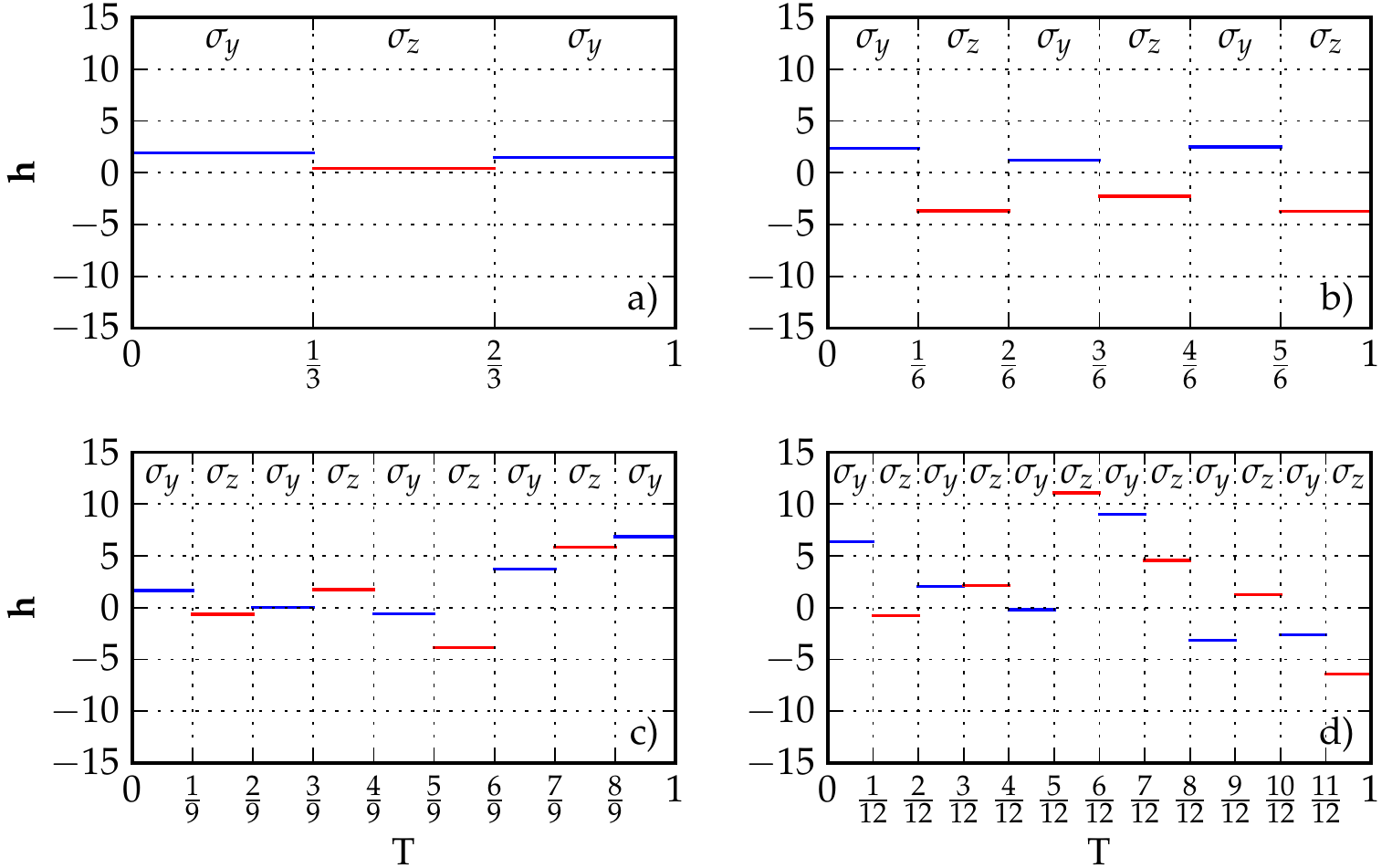}
\caption{Controls that maximize Bob's payoff. Panel a) shows the control vector
that generates Bob's strategy when using 3 control parameters, in this case his
payoff is 0.713. Panels b), c) and d) show Bob's controls of length 6, 9 and 12
giving payoffs 0.804, 0.8241 and 0.8246 respectively.}
\label{fig:controlsBob}
\end{figure}

In Fig.~\ref{fig:controlsBob} the sequences of control pulses allowing Bob to
achieve mean payoff greater than $1/2$ are presented.

%%%%%%%%%%%%%%%%%%%%%%%%%%%%%%%%%%%%%%%%%%%%%%%%%%%%%%%%%%%%%%%%%%%%%%%%%%%%%%%%
\paragraph{Optimization with additional qubits}
%%%%%%%%%%%%%%%%%%%%%%%%%%%%%%%%%%%%%%%%%%%%%%%%%%%%%%%%%%%%%%%%%%%%%%%%%%%%%%%%
Let us now consider the scenario where Bob is able to add more auxiliary qubits
and build a longer spin-chain. Unfortunately, in this situation it is impossible
to find $J/T$ such that the mean payoff for both players using the Pauli
strategy is $1/2$. For this reason we must assume that Bob convinced Alice to
play the qubit flip game using his device without allowing her to check it by
playing with the Pauli strategy.

\begin{table}[ht]
	\centering
	\begin{tabular}{|l||c|c|c|c|c|c|}
		\hline
		Chain length & 2 & 3 & 4 & 5 & 6 & 7 \\\hline
		Maximal payoff & 0.713 & 0.920 & 0.925 & 0.901 & 0.901 & 0.951 \\\hline
	\end{tabular}
	\caption{Bob's maximal payoff for a different size of a chain in the case of
	using three control pulses.}
	\label{tab:bob-different-chains}
\end{table}

The maximal payoffs obtained by Bob in this scenario are presented in
Table~\ref{tab:bob-different-chains}.

% Again, one can consider two methods of choosing strategies by Bob. In the
%first
% case Bob is assuming that the game is fair and he is using the Pauli strategy,
% which would be optimal in the situation when both players have only one qubit
% for their disposal. In the second case, Bob is using random strategy, \ie\ he
% aims to choose his control parameters, so that the resulting evolution is
% random. Note that in both cases, Bob is not aware of the second subsystem and
%he
% calculates his parameters assuming that the system is composed of one qubit
% only.

%%%%%%%%%%%%%%%%%%%%%%%%%%%%%%%%%%%%%%%%%%%%%%%%%%%%%%%%%%%%%%%%%%%%%%%%%%%%%%%%
\section{Concluding remarks}\label{sec:final}
%%%%%%%%%%%%%%%%%%%%%%%%%%%%%%%%%%%%%%%%%%%%%%%%%%%%%%%%%%%%%%%%%%%%%%%%%%%%%%%%
We have discussed some possible realizations of a penny flip game in a~quantum
system. First, we have assumed that both players are aware of the fact that they
play the game using a qubit. In the next step we have extended the model by
adding an additional qubit. 

In the second case only the first player was aware of the fact that the game is
played on a two-qubit system and of the type of interaction between qubits. We have
shown that in such a situation she is able to achieve a mean payoff equal to almost 1.
Moreover, it is enough to use one auxiliary qubit (spin) only to achieve this
effect. This shows that in the situation when one of the parties is not aware
of the dimensionality of the system, the other can easily use this fact to play
unfairly.

To show that also the second player can gain by harnessing the dimensionality 
of the system we have introduced a scenario where he tries to cheat against the 
first player. We have shown that the payoff of the second
player grows with the number of controls and with the number of auxiliary qubits.

The presented work leaves open a problem of the behaviour of the game in the
situation where players, or at least one of them, are able to use more general
strategies, namely strategies described by quantum channels. This would also
allow to incorporate the influence of noise on the course of the proposed
game.

\begin{acknowledgements}
Work by P.~Gawron was supported by the Polish Ministry of Science and Higher
Education under the grant number IP2010~009770, J.A.~Miszczak was
supported by Polish National Science Centre under the project number 
N~N516~475440,
Z.~Pucha{\l}a was supported by the Polish National Science Centre under 
the research project N~N514~513340. %Z.P. iuventus = IP2010 033470.
\end{acknowledgements}

\bibliography{qubit_flip}
\bibliographystyle{spmpsci}
\end{document}